\documentclass[12pt]{iopart}
\usepackage{epsfig}

\newcommand{\ANP}{{\it Adv. Nucl. Phys.} }
\newcommand{\EPJ}{{\it Eur. Phys. J.} }

\newcommand{\PRP}{{\it Phys. Rep.} }

\newcommand{\beq}{\begin{equation}}
\newcommand{\eeq}{\end{equation}}
\newcommand{\bea}{\begin{eqnarray}}
\newcommand{\eea}{\end{eqnarray}}
\newcommand{\bce}{\begin{center}}
\newcommand{\ece}{\end{center}}

\def\lsim{\mathrel{\rlap{\lower4pt\hbox{\hskip1pt$\sim$}}
    \raise1pt\hbox{$<$}}}         
\def\gsim{\mathrel{\rlap{\lower4pt\hbox{\hskip1pt$\sim$}}
    \raise1pt\hbox{$>$}}}         

\begin{document}

\title{Theory and Phenomenology of Vector Mesons in Medium} 

\author{Ralf Rapp\footnote[3]{email: rapp@comp.tamu.edu} 
}

\address{Cyclotron Institute and Physics Department, Texas A\&M University, 
               College Station, Texas 77843-3366, U.S.A.}

\begin{abstract}
Electromagnetic probes promise to be direct messengers of (spectral 
properties of) hot and dense matter formed in heavy-ion collisions, 
even at soft momentum transfers essential for characterizing possible 
phase transitions. We examine how far we have progressed toward this 
goal by highlighting recent developments, and trying to establish 
connections between lattice QCD, effective hadronic models and 
phenomenology of dilepton production.  
\end{abstract}




\section{Introduction}
\label{sec_intro}
High-energy heavy-ion physics aims at investigating the properties
of strongly in\-teracting matter under extreme conditions, at much higher
temperatures ($T$) and baryon densities ($\varrho_B$) than accessible
in atomic nuclei. A major goal is to establish the existence
of new states of matter characterized by the restoration of the
spontaneously broken chiral symmetry and the deconfinement of color
charges, as predicted by Quantum Chromodynamics (QCD). Hadronic
observables are widely (and successfully) used to infer the 
composition and collective expansion of the highly excited system 
and thus determine its bulk properties, i.e., equation of state (EoS). 
Direct information on the
degrees of freedom and spectral properties of the matter are 
harder to obtain, especially at soft scales typical for the phase 
transition, $\sim$$T_c$, $\Lambda_{\rm QCD}$,  up to the chiral 
scale $\Lambda_\chi$$\sim$$4\pi f_\pi$ ($f_\pi$=92~MeV: pion decay 
constant). Electromagnetic (e.m.) probes (real and 
virtual photons) provide a unique opportunity to study the $J^P=1^-$
response of the strongly interacting 
medium~\cite{Rapp:1999ej,Alam:1999sc,Cassing:1999es,Gale:2005zd}, 
defined by the pertinent retarded correlation function, 
\begin{equation}
\Pi_{\rm em}^{\mu\nu}(q;\mu_B,T) = -i \int d^4x \ \Theta(t) \ 
{\rm e}^{iqx} \ 
\langle\Omega| j^\mu_{\rm em}(x) j^\nu_{\rm em}(0)|\Omega \rangle  
\label{Pi_em}
\end{equation}    
($|\Omega \rangle$: finite $T,\mu_B$ ground state).
From $e^+e^-$$\to$$hadrons$ one knows that, in the vacuum, the low-mass 
($M$$\lsim$$\,\Lambda_\chi$) e.m.~spectral function is saturated by 
light vector mesons, 
\begin{equation}
{\rm Im}\Pi_{\rm em} \sim \left[ {\rm Im} D_\rho  + 
\frac{1}{10}{\rm Im} D_\omega + \frac{1}{5} {\rm Im} D_\phi \right] \ ,   
\label{ImPi_em}
\end{equation}
which also illustrates the dominant role of the $\rho$. In 
vector-dominance models, medium modifications are thus encoded in the 
vector-meson propagators, $D_V$ ($V =\rho, \omega, \phi$).

The study of vector mesons in hadronic matter should be placed in a 
broader context. This includes questions of the type:
(1) Do $\rho$, $\omega$ and $\phi$ mesons behave alike?
(2) Are hadronic interactions sufficient to account for medium effects 
    or are ``intrinsic" $T$- and $\mu_B$-dependencies of the parameters 
    of the effective Lagrangian to be 
     included~\cite{Harada:2003jx,Brown:2002is}?
(3) How do vector-meson spectral functions change across the phase
    diagram (i.e., baryon- vs. meson-induced modifications)?
(4) What is the impact of medium-modified hadron properties 
    on the EoS and its chemical composition?
(5) What are the connections between vector spectral functions
    and (chiral) order parameters (quark and gluon condensates, $f_\pi$, 
    susceptibilities, ...), and how can they be exploited?    
The nonperturbative nature of these questions requires the use of
effective (chiral) models with careful constraints from both first 
principles (symmetries, lattice QCD) and experiment. Phenomenological 
applications to electromagnetic observables in heavy-ion collisions 
can then pave the way to deduce the onset and realization of chiral
symmetry restoration, a key prediction of QCD.    

In this paper, we discuss 3 basic elements (lattice QCD, effective 
models and pheno\-menology) to address the above issues. 
Sec.~\ref{sec_lqcd} recalls relations between e.m.~spectral
functions and chiral order parameters via 
susceptibilities and sum rules.
Sec.~\ref{sec_had} gives a brief survey on vector mesons in 
hot/dense hadronic matter. 
Sec.~\ref{sec_dilep} assesses ingredients for studying dilepton
spectra in heavy-ion collisions, with applications to, and 
interpretations of, recent SPS data\footnote{Medium effects on vector 
mesons in elementary 
(photo-/electro-/hadro-) production off nuclei have been discussed 
in excellent review talks by Djalali~\cite{Djalali:2007} and 
Metag~\cite{Metag:2007} and will not be reiterated here.}.
Sec.~\ref{sec_concl} contains concluding remarks.

\section{Vector Correlators from Lattice QCD and Sum Rules}
\label{sec_lqcd}
\begin{figure}[!tb]
\center
\hspace{0.8cm}
\begin{minipage}{7cm}
\epsfig{file=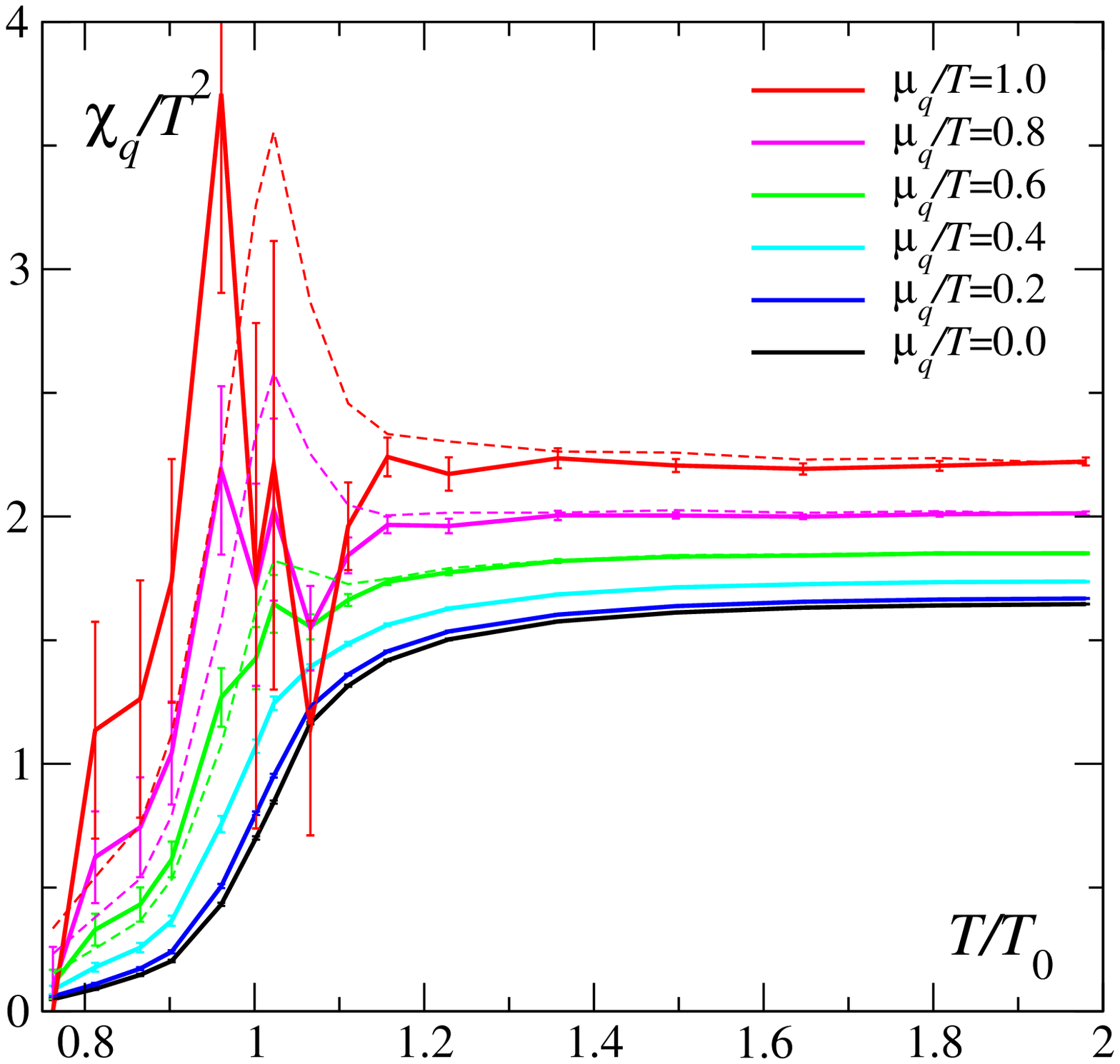,width=6cm}
\end{minipage}
\hspace{0.5cm}
\begin{minipage}{7cm}
\epsfig{file=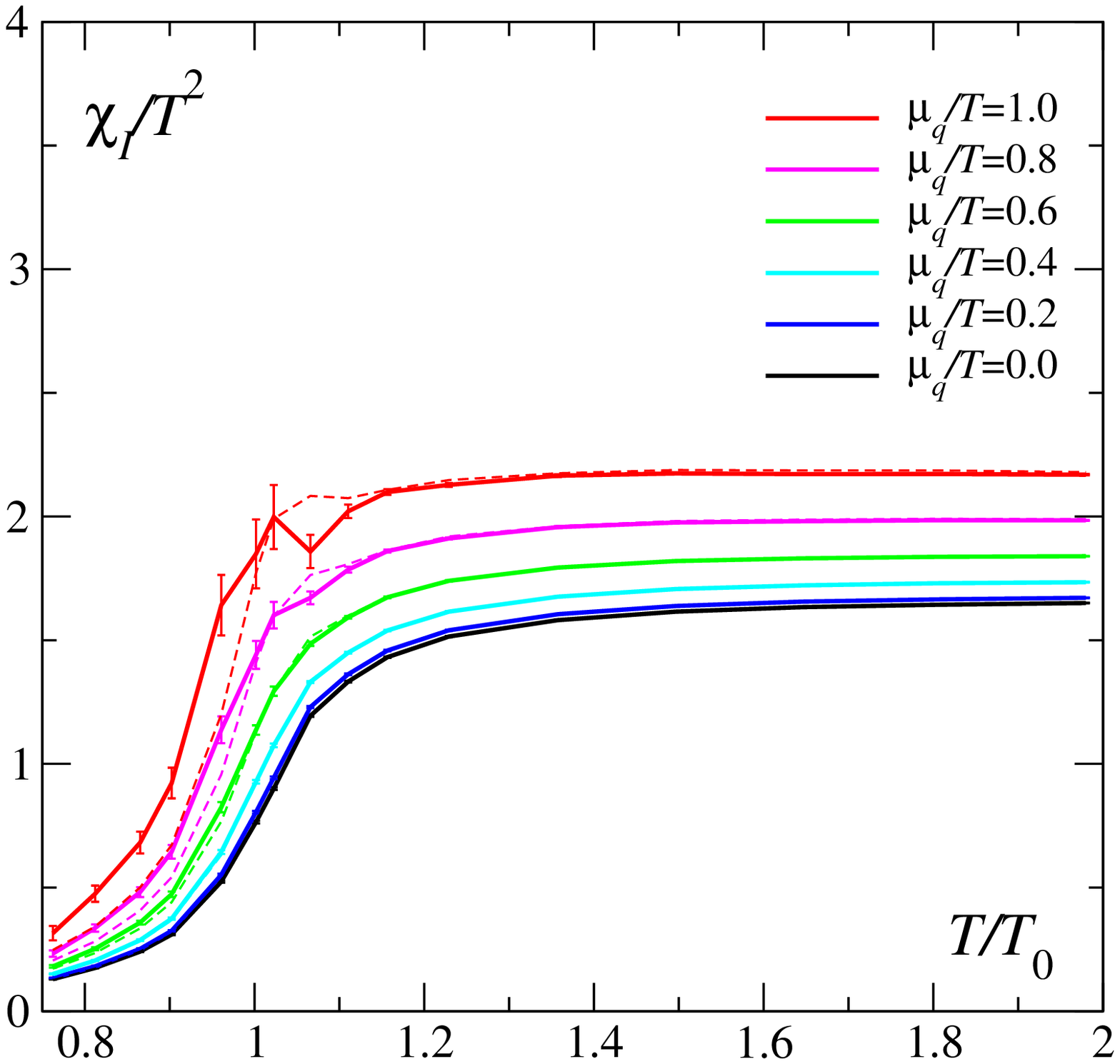,width=6cm}
\end{minipage}
\caption{Isoscalar (left) and isovector (right) quark-number 
susceptibility for various quark chemical potentials, 
$\mu_q=\mu_B/3$, as computed in unquenched 
lattice QCD~\cite{Allton:2005gk}.}
\label{fig_chi-lat}
\end{figure}

The basic quantity which, in principle, contains all the information
about an interacting system at finite $T$ and $\mu_B$ (baryon chemical
potential) is the thermodynamic potential, 
\begin{equation}
\Omega(\mu_B,T) = -T \log {\cal Z}  = -P V ,  
\end{equation}
defined in terms of the partition function, ${\cal Z}$ ($P$: pressure).
Its $T$ and $\mu_B$ derivatives give rise to the EoS (entropy, 
energy-/density), while hadronic correlation functions follow 
from expectation values of suitable currents, cf.~Eq.~(\ref{Pi_em}).
The EoS is routinely computed in lQCD, but the formulation of lQCD
in Euclidean (imaginary) time renders it more difficult to 
extract spectral functions in the timelike regime.  
A valuable interface of bulk and microscopic quantities is provided 
by susceptibilities, which are 2.~order derivatives of $\Omega$ that 
can be related to the spacelike (screening) limit of pertinent 
correlators,
\begin{equation}
\chi_\alpha \sim \frac{\partial^2 \Omega}{\partial \mu_\alpha^2} \sim 
\Pi_\alpha(q_0=0,q\to 0) \ .
\end{equation}
Fig.~\ref{fig_chi-lat} shows a recent lQCD computation of the 
$T$-dependence of light quark-number susceptibilities in the isoscalar 
($\alpha$=$q$) and isovector ($\alpha$=$I$) channel, carrying the 
quantum number of the $\omega$ and $\rho$ meson, respectively.
With increasing $\mu_q$ (evaluated via a Taylor expansion), the 
isovector channel behaves smoothly while the isoscalar one develops a 
pronounced peak. This could be related to the approach to a critical 
endpoint in the phase diagram, possibly induced by  
the effect of a $\sigma$-$\omega$ mixing. 

A powerful tool to relate chiral order parameters to (in-medium) 
$V$-meson spectral functions are sum rules (SRs). In QCD-SRs a 
(subtracted) dispersion relation for a correlation function is 
formulated for spacelike momenta $q^2$=$-Q^2$$<$0~\cite{Shifman:1978bx}, 
\begin{equation}    
\Pi_\alpha(Q^2) = \Pi_\alpha(0) +c Q^2 
+ Q^4 \int \frac{ds}{s^2} \frac{\rm Im \Pi_\alpha(s)}{s+Q^2} \ .   
\label{qcdsr}
\end{equation} 
The spectral function on the right-hand-side is related to quark and
gluon condensates which arise as nonperturbative coefficients within
an operator-product expansion (OPE) in 1/$Q^2$ of $\Pi_\alpha(Q^2)$ 
on the left-hand-side. The condensates on the OPE side generally  
decrease in the medium, which, in turn, requires a softening of the 
spectral function under the dispersion integral. The softening can 
be satisfied by both broadening and a downward mass 
shift~\cite{Leupold:1997dg,Ruppert:2005id}. For $\omega$ and $\rho$ mesons 
the OPE side is very similar (governed by 4-quark condensates), 
while the subtraction constant differs considerably, 
$\Pi_\rho(0)=\Pi_\omega(0)/9$, implying stronger 
medium effects (softening) on the $\rho$ than on the $\omega$.

For the vector-isovector ($\rho$) channel (which dominates the dilepton 
rate, cf.~eq.~(\ref{ImPi_em})), so-called Weinberg and DMO sum
rules~\cite{Weinberg:1967,Das:1967} are of special interest. 
Using current algebra, one can relate moments of the difference 
between vector and 
axialvector spectral functions to chiral order parameters. In the
chiral limit ($m_\pi$=0) one has  
\begin{eqnarray}
f_n = - \int\limits_0^\infty \frac{ds}{\pi} \ s^n \
 \left[{\rm Im} \Pi_V(s) - {\rm Im} \Pi_A(s) \right]  \ ,
\qquad \qquad \qquad
\label{csr}
\\
f_{-2} = f_\pi^2  \frac{\langle r_\pi^2 \rangle}{3} - F_A \ , \quad
f_{-1} = f_\pi^2 \ , \quad
 f_0   = 0 \ ,  \quad
f_1 = -2\pi \alpha_s \langle {\cal O} \rangle  \ 
\label{fn}
\end{eqnarray}
($r_\pi$: pion charge radius, $F_A$: radiative pion decay constant, 
$\langle {\cal O} \rangle$: 4-quark condensate). Weinberg sum rules
remain valid at finite $T$~\cite{Kapusta:1993hq}, demonstrating that 
chiral restoration requires degeneracy of the entire spectral functions. 
Combining lQCD computations of order parameters with effective model 
calculations of spectral functions thus provides a very promising 
synergy for deducing chiral restoration from experiment~\cite{David:2006sr}.

\section{Hadronic Spectral Functions}
\label{sec_had}
\begin{figure}[!tb]
\begin{minipage}{7.7cm}
\epsfig{file=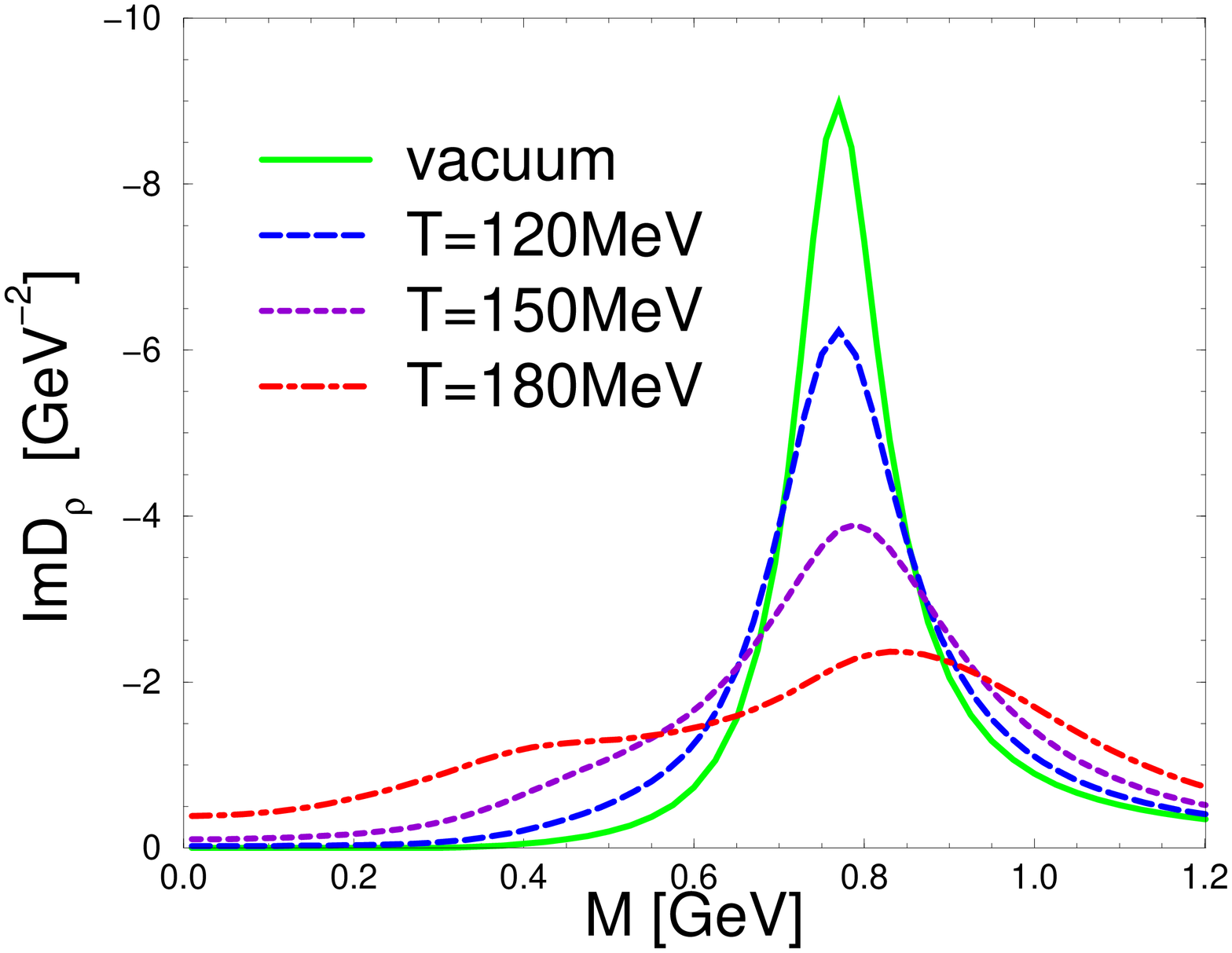,angle=-90,width=\linewidth}
\end{minipage}
\hspace{-0.5cm}
\begin{minipage}{7.7cm}
\epsfig{file=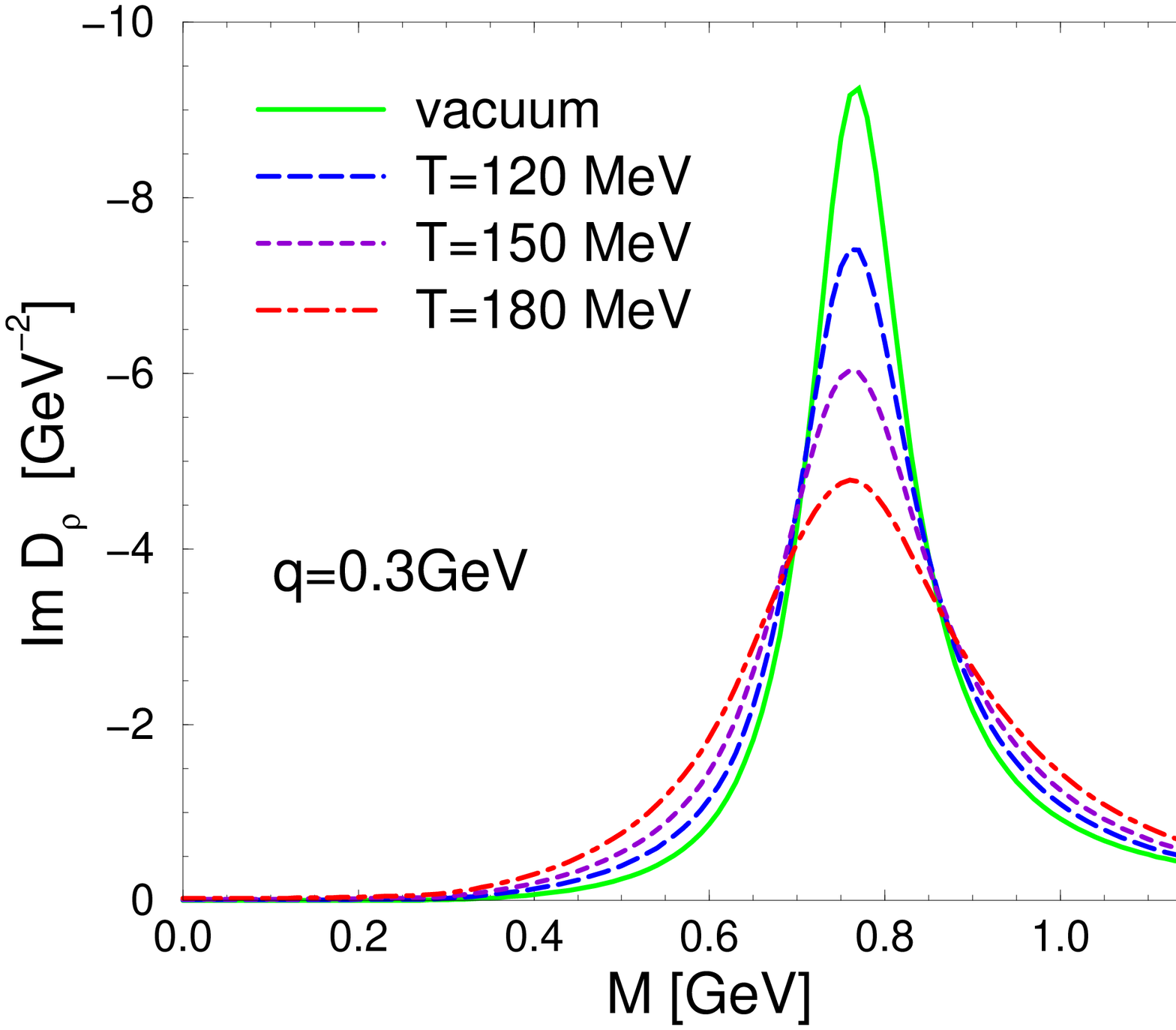,width=\linewidth}
\end{minipage}
\caption{$\rho$-meson spectral function in a hadronic many-body
approach. Left panel: hot hadronic matter~\cite{Rapp:1999us} 
at fixed $\mu_B$=330~MeV (approximately reflecting CERN-SPS conditions) 
corresponding to 
($\varrho_B$/$\varrho_0$,$T$/MeV)$\simeq$(0.1,120),(0.7,150),(2.6,180);
right panel: hot meson gas (no anti-/baryons)~\cite{Rapp:1999qu}. 
}
\label{fig_Arho69}
\end{figure}

A commonly used approach to evaluate in-medium $V$-meson spectral 
functions is hadronic many-body theory~\cite{Rapp:1999ej}. The key 
quantity figuring into the $V$-meson propagator, 
\begin{equation}
D_V(M,q;\mu_B,T) = \left[M^2-m_V^2-\Sigma_{VP}-\Sigma_{VB}
-\Sigma_{VM}\right]^{-1} \ ,
\end{equation}
is the selfenergy, $\Sigma_{V}$, which can be calculated from 
effective interactions integrated 
over the momentum distributions of mesons ($M$=$\pi$,$K$,$\rho$,...) 
and baryons ($B$=$N$,$\Lambda$,$\Delta$,...) in the heat bath 
($\Sigma_{VP}$ encodes medium-modified $\pi\pi$, $3\pi$, $K\bar K$ 
decays). The interaction vertices ought to be compatible with gauge 
and chiral invariance and constrained by resonance decay widths and 
scattering data in the vacuum. Typical results for the $\rho$
spectral function are 
shown in Fig.~\ref{fig_Arho69}. 
Prominent features are a strong broadening with little mass 
shift and the prevalence of anti-/baryonic effects, especially below 
the free $\rho$ mass.
When extrapolated to the expected phase boundary, the $\rho$ spectral
function ``melts" and resembles the form and magnitude
of the perturbative $q\bar q$ correlator. In cold nuclear matter, 
it is compatible with QCD sum 
rule analysis~\cite{Leupold:1997dg} and qualitatively consistent 
with the new CLAS data for photoproduction 
off nuclei at JLab~\cite{Djalali:2007}.  

\begin{figure}[!tb]
\begin{minipage}{7.5cm}
\epsfig{file=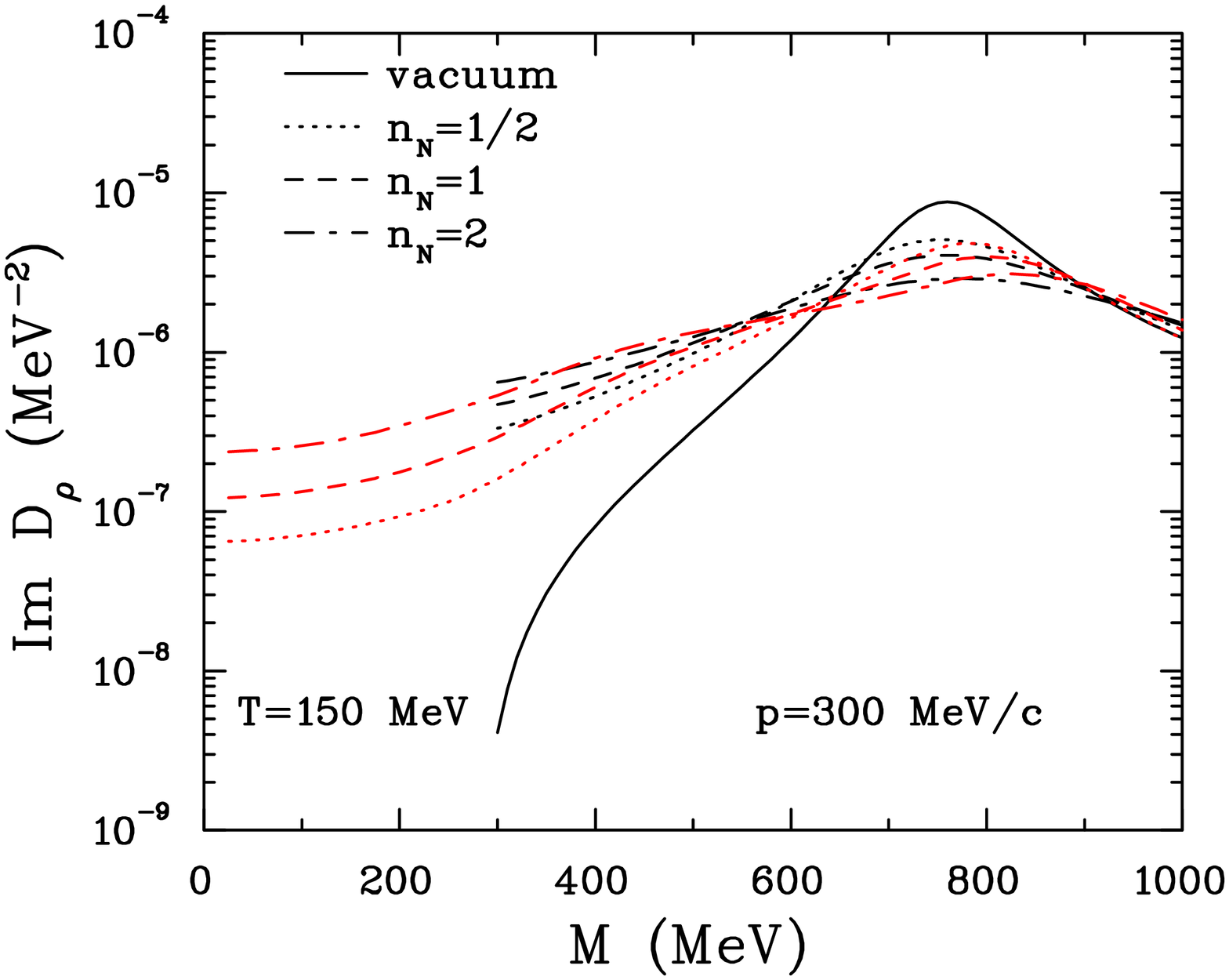,angle=90,width=\linewidth}
\end{minipage}
\hspace{0.5cm}
\begin{minipage}{7.5cm}
\epsfig{file=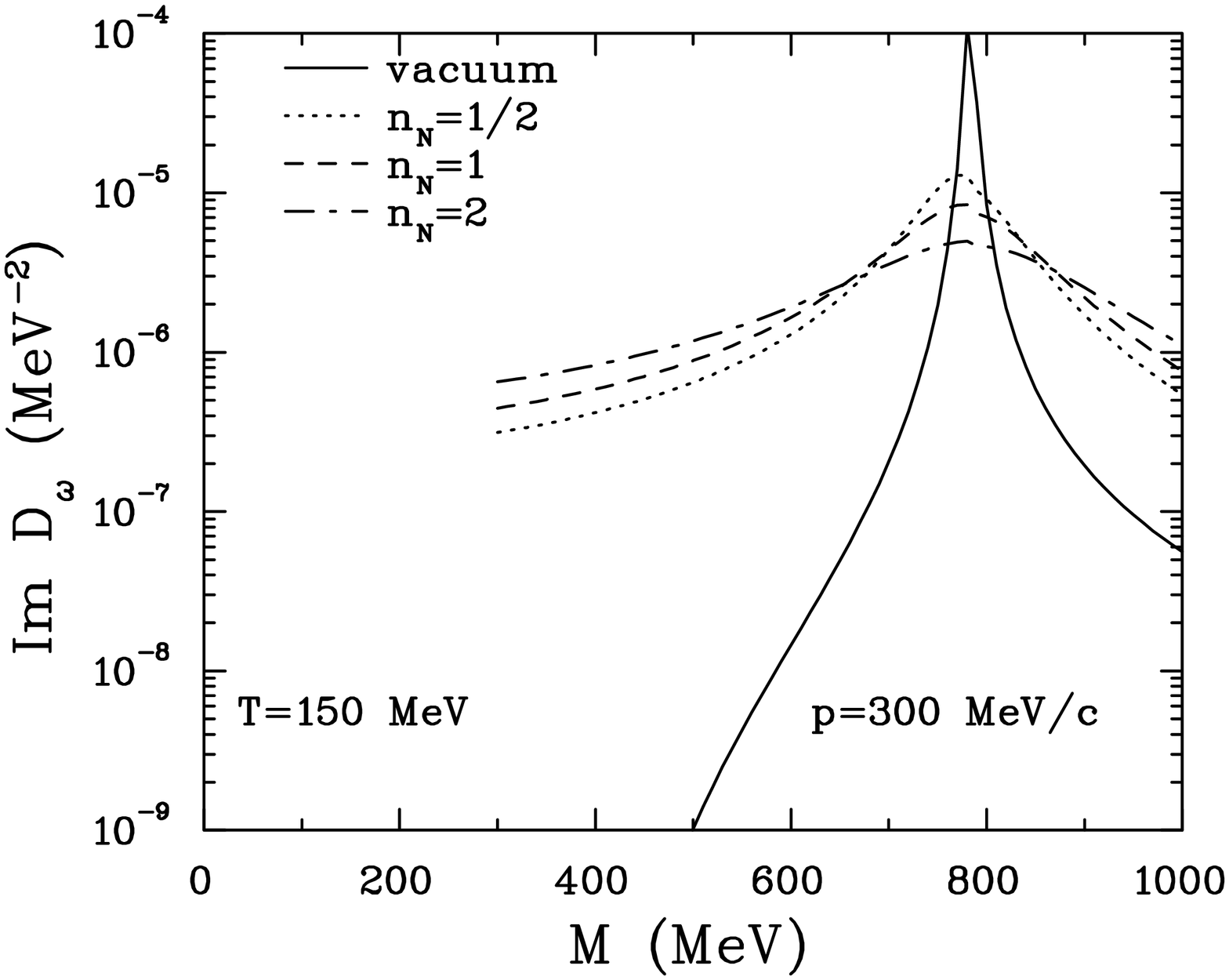,angle=90,width=\linewidth}
\end{minipage}
\caption{$V$-meson spectral functions in a hot $\pi$-$N$ gas based 
on empirical scattering amplitudes constrained by resonances and Regge 
behavior~\cite{Eletsky:2001bb}.   
Left panel: $\rho$ meson, compared to hadronic many-body
calculations of Fig.~\ref{fig_Arho69}.
Right panel: $\omega$ meson.
}
\label{fig_Arho-elets}
\end{figure}

In Ref.~\cite{Eletsky:2001bb}, the $\Sigma_V$'s have been 
evaluated in terms of empirical scattering amplitudes as constructed 
from resonance dominance at low energies and Regge-type behavior at high
energies, while the real parts have been calculated using dispersion
relations. It is satisfying to see that the resulting $\rho$ spectral
function agrees quite well with the many-body approach, cf.~left
panel of Fig.~\ref{fig_Arho-elets} (the largest discrepancy emerges 
toward the 2$\pi$ threshold; also, some care has to be taken in 
identifying nucleon and baryon densities at finite $T$). 
The $\omega$ spectral function (right panel of 
Fig.~\ref{fig_Arho-elets}) also broadens considerably (with
little mass shift), but less than the $\rho$, which is in line
with the expectation from QCD sum rules (recall the discussion
after eq.~(\ref{qcdsr})).

Within the Hidden Local Symmetry approach, it has been suggested to 
identify the chiral partner of the (longitudinal) $\rho$ with the $\pi$ 
(rather than $a_1$)~\cite{Harada:2003jx}. When applied within a 
finite-$T$ loop expansion, and after matching the hadronic
axial-/vector correlators to pQCD in the spacelike regime, one
finds a dropping $\rho$ mass as well as a violation of vector dominance.
The finite-$T$ e.m.~formfactor, which enters into the dilepton
production rate, clearly exhibits the downward moving $\rho$ 
peak~\cite{Harada:2006hu}. It would be interesting to see whether 
this feature persists when including the effects of baryons. 

\section{Dilepton Phenomenology in Heavy-Ion Collisions}
\label{sec_dilep}
The dilepton emission rate is directly proportional to the
e.m.~spectral function~\cite{MT85}, 
\begin{equation}
\frac{dN_{ll}}{d^4xd^4q} = -\frac{\alpha_{\rm em}^2}{\pi^3 M^2} \
       f^B(q_0;T) \  {\rm Im}\Pi_{\rm em}(M,q;\mu_B,T) \ .
\label{Rll}
\end{equation}
Assuming the medium in a heavy-ion collision to kinetically equilibrate,
thermal dilepton spectra can be calculated by convolution over
a realistic space-time model. Obviously, the latter must be consistent 
with hadronic chemistry (evolution of $\mu_B$ and $T$) and $p_T$ 
spectra (collective expansion). However, depending on centrality and
invariant mass or $q_T$ of the lepton pair, nonthermal effects/sources, 
such as Drell-Yan annihilation and final-state decays 
(including correlated open-charm), may not be neglected.

\begin{figure}[!tb]
\begin{minipage}{7.6cm}
\vspace{-0.2cm}
\epsfig{file=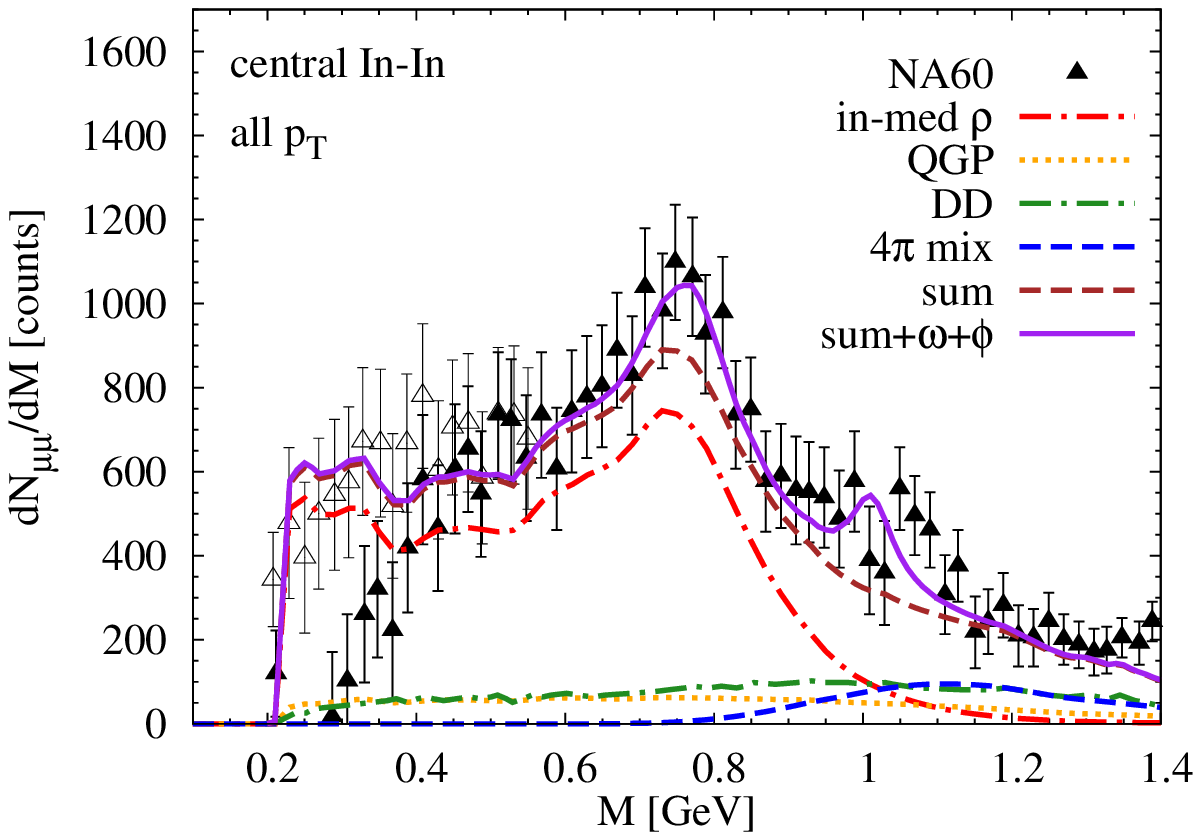,width=1.0\linewidth}
\end{minipage}
\hspace{0.2cm}
\begin{minipage}{7.6cm}
\epsfig{file=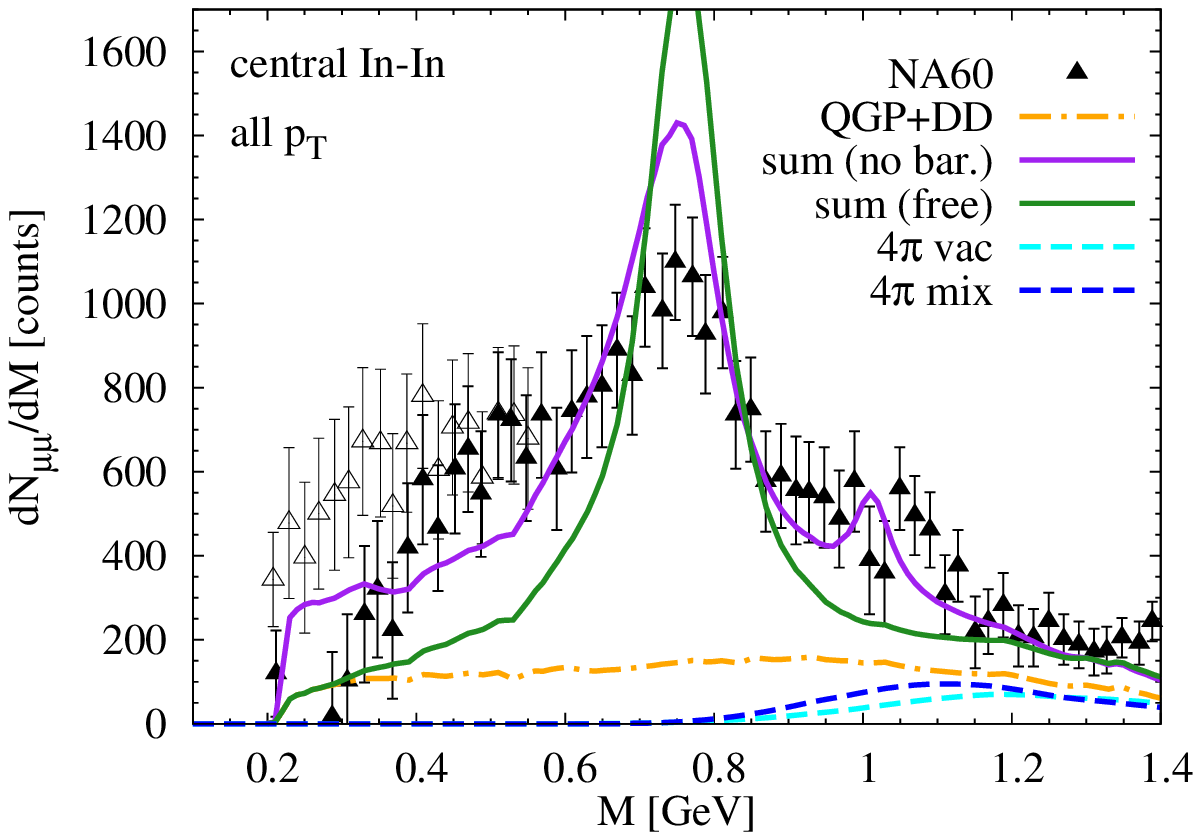,width=1.0\linewidth}
\end{minipage}
\begin{minipage}{7.6cm}
\vspace{-0.2cm}
\epsfig{file=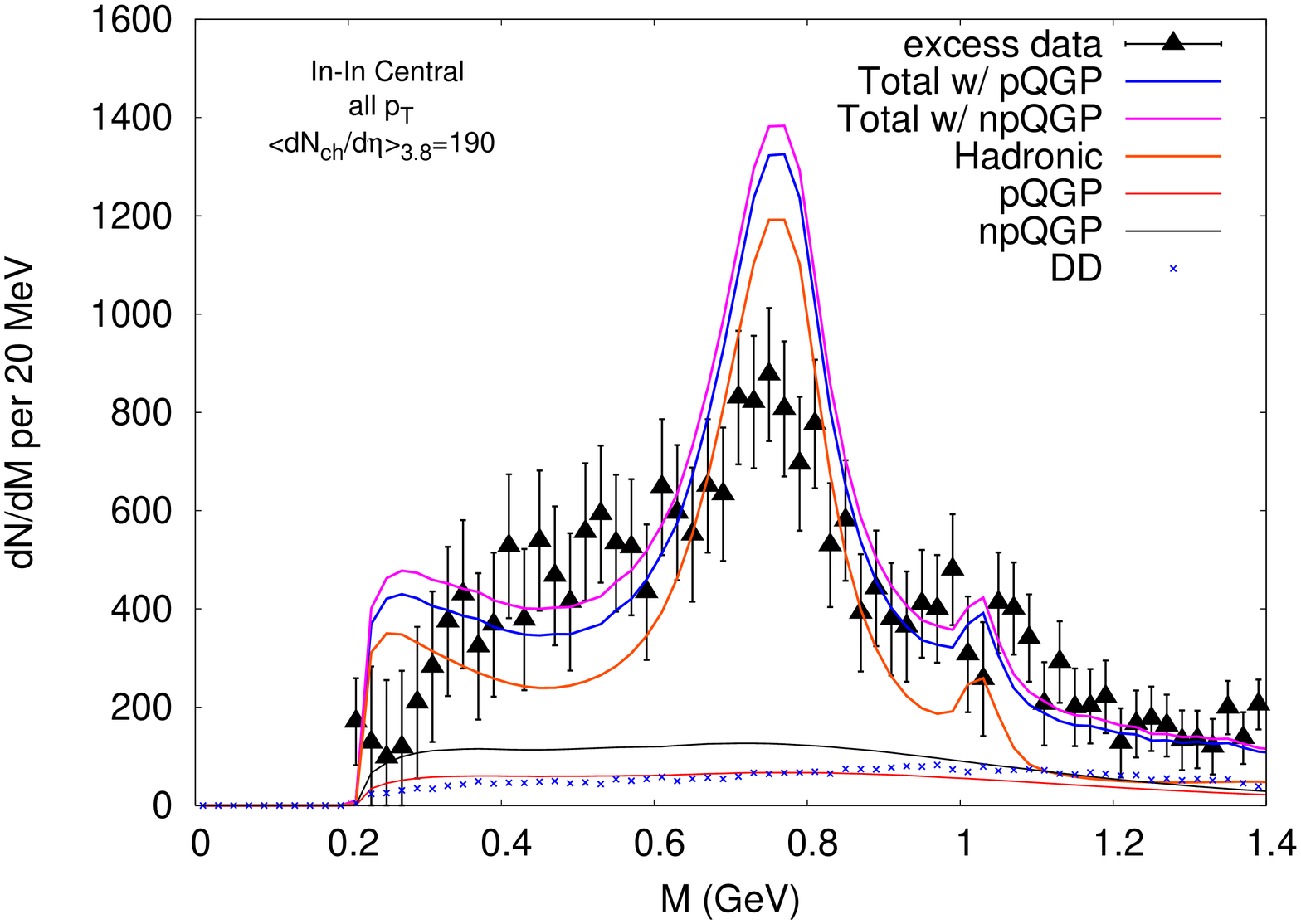,width=1.02\linewidth}
\end{minipage}
\hspace{0.4cm}
\begin{minipage}{7.6cm}
\epsfig{file=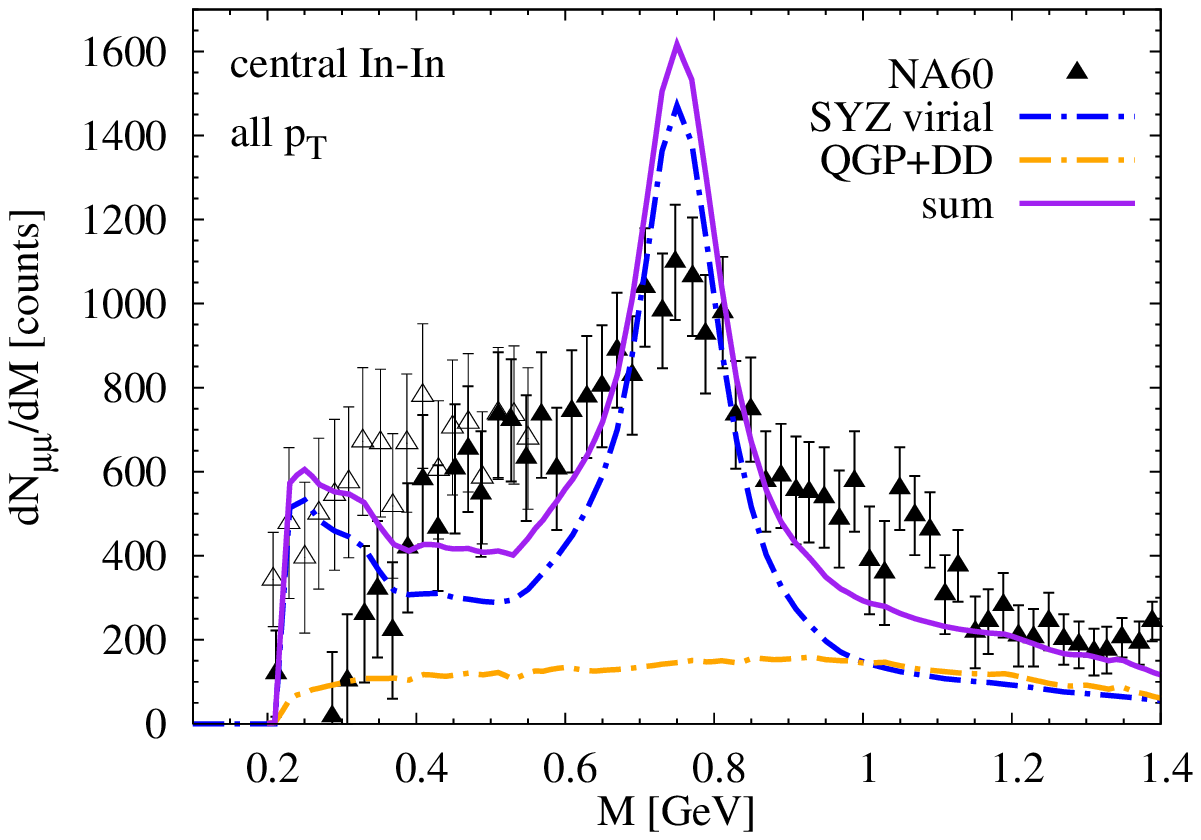,width=0.98\linewidth}
\end{minipage}
\caption{NA60 dimuon invariant-mass spectra in central 158~AGeV In-In
collisions~\cite{Arnaldi:2006jq}, compared to thermal fireball emission
based on hadronic many-body theory and $q\bar q$ annihilation in a 
QGP~\cite{vanHees:2006ng}  
(upper panel; left: full results, right: without baryon effects and 
when using the free e.m.~spectral function). Lower panels: thermal 
spectra using e.m.~rates from a chiral virial expansion within a 
hydrodynamic evolution~\cite{Dusling:2006yv} (left) and a thermal 
fireball~\cite{vanHees:2006ng} (right). 
}
\label{fig_na60-m}
\end{figure}

In the upper panels of Fig.~\ref{fig_na60-m} thermal fireball 
calculations employing $V$-meson spectral functions from hadronic
many-body theory~\cite{vanHees:2006ng} are compared to recent NA60 data 
(where improved experimental accuracy enabled a subtraction of 
final-state hadron decay contributions, except for $\rho$ and charm 
decays)~\cite{Arnaldi:2006jq}. The low-mass region is well described 
by the in-medium spectral functions (upper left panel), supporting 
the predicted $\rho$ line shape (QGP emission is rather small). 
When switching off baryonic effects (or all medium effects), the 
spectra are incompatible with the data (upper right panel).
In fact, the space-time integrated dilepton line shapes for different
scenarios reflect the changes of the $\rho$ spectral function 
in Fig.~\ref{fig_Arho69}, confirming good sensitivity to medium effects. 
At masses above 1~GeV, the free e.m.~correlator mostly couples to 4-pion
states. Inclusion of this contribution in the fireball emission leads 
to reasonable agreement with the observed enhancement at $M$$\ge$1~GeV 
(effects of leading-$T$ vector-axialvector mixing~\cite{Dey:1990ba} 
increases the 4$\pi$ yield slightly). 
Note that the relative strength of the various thermal sources is fixed; 
the absolute yield is sensitive to the fireball lifetime, which in
Fig.~\ref{fig_na60-m} amounts to about 7~fm/c (the initial/final 
temperatures are $T_{\rm i,fo}$$\simeq$195,120~MeV).  
The lower panels of Fig.~\ref{fig_na60-m} display results of using
the in-medium e.m.~spectral function calculated in a 3-flavor chiral
reduction formalism~\cite{Steele:1997tv}, which couples chiral Ward
identities with a low-density expansion using empirical
$\rho$-/$\gamma$-$N$ and $\rho$-$\pi$ scattering amplitudes. The low-
and intermediate-mass enhancement is well described, but the lack of
$\rho$-broadening (inherent to the low-density expansion) leads to an
overestimate around the free $\rho$ mass. The spectra obtained from
the hydrodynamic evolution used in Ref.~\cite{Dusling:2006yv} (left
panel) agree well with the fireball approach of
Ref.~\cite{vanHees:2006ng} (right panel). This also holds for the
(smallness of the) perturbative QGP contribution. On the contrary,
in Ref.~\cite{Renk:2006qr} QGP radiation is the dominant source
above $M\simeq0.9$~GeV.

\begin{figure}[!tb]
\begin{minipage}{7.5cm}
\epsfig{file=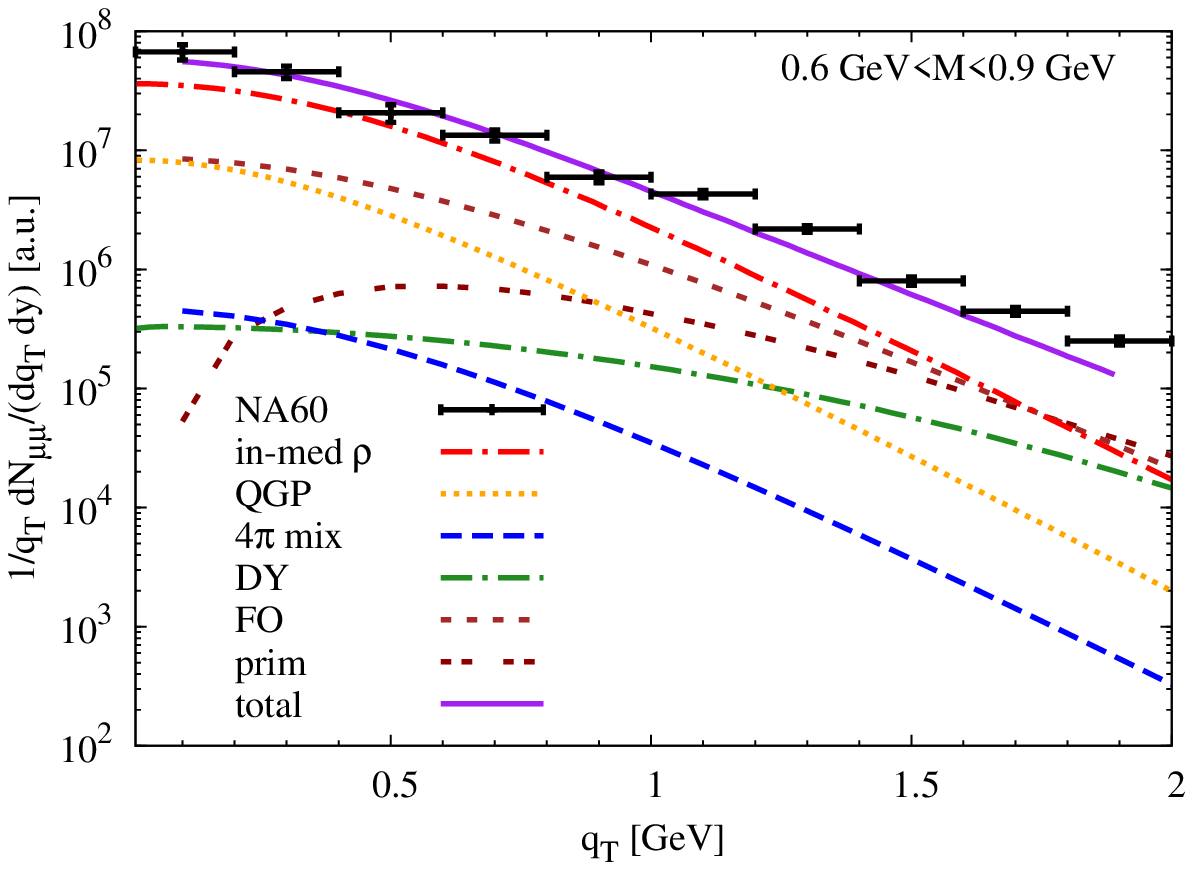,width=0.98\linewidth}
\end{minipage}
\hspace{0.2cm}
\begin{minipage}{7.5cm}
\epsfig{file=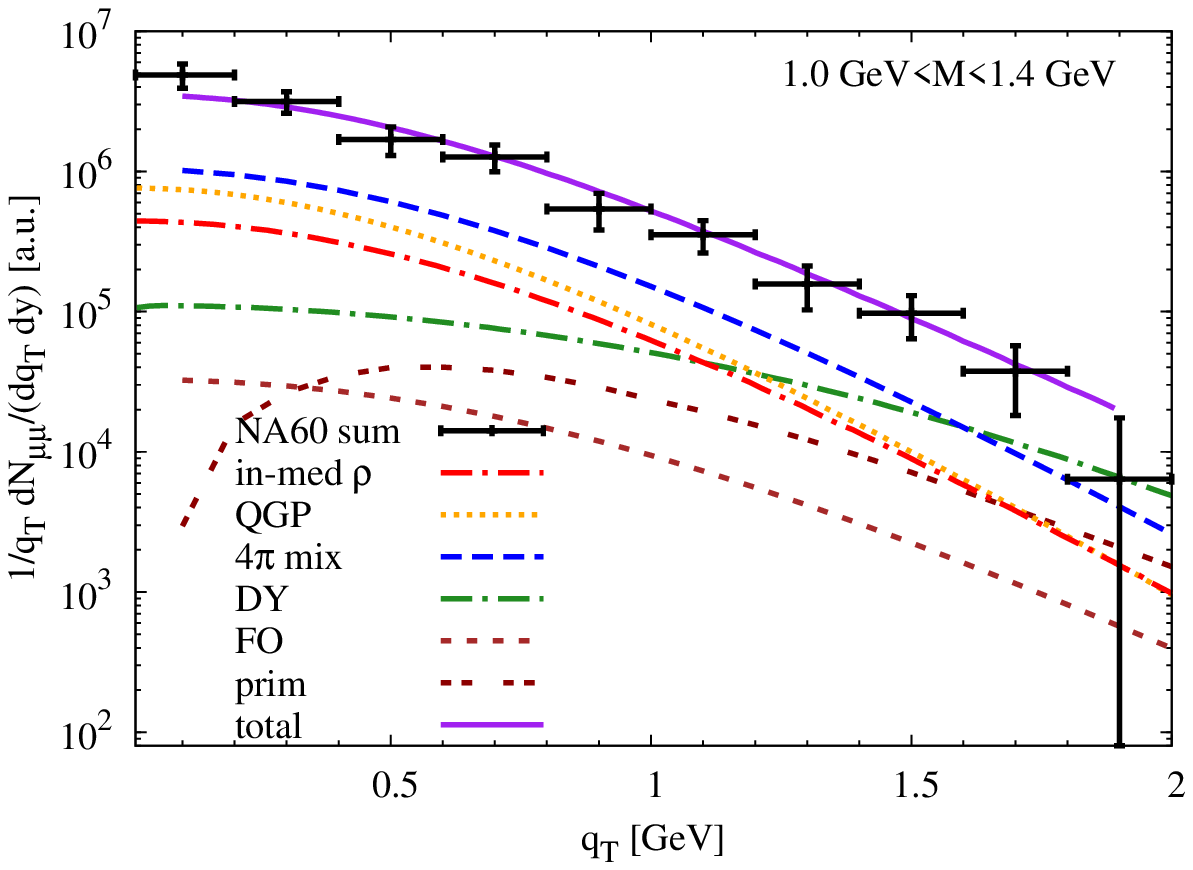,width=1.03\linewidth}
\end{minipage}
\caption{Dimuon $q_T$ spectra in semicentral 158~AGeV  
In-In collisions~\cite{Damjanovic:2007qm} for mass bins
0.6~GeV$<$$M$$<$0.9~GeV (left) and 1.0~GeV$<$$M$$<$1.4~GeV 
(right), compared to thermal fireball calculations with in-medium 
$\rho$, $\omega$, $\phi$ spectral functions, $4\pi$ annihilation,
QGP emission, thermal-freezeout and primordial $\rho$'s, as well as 
Drell-Yan annihilation~\cite{vanHees:2007xx}.
}
\label{fig_na60-pt}
\end{figure}

Additional information on the nature of the excess radiation in 
heavy-ion collisions can be obtained from pair-$q_T$ spectra, especially
when differential in mass, see Fig.~\ref{fig_na60-pt}.
At low $q_T$ the spectra are dominated by thermal radiation (either
$2\pi$ or $4\pi$ annihilation), but the composition becomes more 
involved at higher $q_T$. In the $\rho$-mass region,  
$\rho$ decays at freezeout (subject to a large flow-induced
blueshift, with in-medium spectral shape) as well as primordial
$\rho$'s from minijets escaping the medium (including a Cronin effect,
with vacuum spectral shape) become significant. Note that the 
emission kinematics of these sources actually differs from thermal 
radiation by a time dilation factor $\gamma$=$q_0/M$. In the $\rho$-mass
bin, the theoretical $q_T$ spectra appear somewhat too soft (as more
clearly revealed in a local slope analysis~\cite{Damjanovic:2007qm}),
cf.~also Ref.~\cite{Renk:2006qr}.
An interesting possibility is that viscosity effects in the later 
hadronic stage render hadron spectra harder~\cite{Teaney:2007}.     
Transverse-momentum and centrality dependencies certainly deserve 
further study.

The CERES/NA45 collaboration has recently published dielectron excess
spectra for central Pb(158~AGeV)+Au~\cite{Adamova:2006nu}. 
Albeit inferior in statistics to the NA60 dimuons, the small 
electron mass allows access to the mass region below the 2$\pi$ 
threshold, where theory predicts a large enhancement 
(the data might show first indications of this). 
In addition, in larger collision systems, the relative size 
of nonthermal contributions is suppressed. Hadronic many-body 
calculations describe the CERES spectra well.

\section{Conclusions}
\label{sec_concl}
E.m.~probes are progressively improving our understanding of microscopic 
properties of hot and dense matter. In this paper we have focused 
on soft emission, which predominantly relates to the physics 
of in-medium vector mesons.
Hadronic models, when carefully constrained, predict a strong in-medium
broadening (``melting") of the $\rho$ resonance which has withstood 
a substantial increase in accuracy in recent dilepton measurements at 
the SPS. This also corroborates the notion 
of (anti-) baryons as the major agents of medium effects, which is
expected to hold even at RHIC and LHC, raising intriguing questions
on the nature of chiral restoration. The determination of the
in-medium axialvector spectral function remains a pressing issue. 
It can and will be addressed by constructing realistic chiral models 
in a hot {\it and} dense medium, which will serve as a bridge between 
order parameters (as computed, e.g., in lattice QCD) and experimental 
data, by utilizing chiral and QCD sum rules.


\vspace{0.5cm}

\noindent
{\bf Acknowledgment} \\
I thank the conference organizers for the 
invitation to an exciting meeting, and H.~van Hees for fruitful 
collaboration. 
This work was supported in part by a U.S.~National Science Foundation
CAREER award under grant no. PHY-0449489.

\section*{References}

\end{document}